\documentclass[letterpaper,arxiv]{dae-handout}
\graphicspath{{./}{./FIGURES/}{./figures/}{../Figures/}{../../Figures/}{./images/}{./graphics/}}

\newcommand{\trversion}{v0.1}
\newcommand{\trfilename}{Markovianity-of-Time.tex}
\newcommand{\trdate}{02026-MAR-13}
\newcommand{\trshorttitle}{The Markovianity of Time}
\newcommand{\trauthor}{Paul Borrill}
\newcommand{\traffiliation}{D\AE D\AE LUS}

\usepackage[T1]{fontenc}
\usepackage[utf8]{inputenc}
\IfFileExists{mdbch.sty}{\usepackage[bitstream-charter]{mathdesign}}{}
\usepackage{amsmath}
\usepackage{amsthm}
\usepackage{mathtools}
\usepackage{booktabs}
\usepackage{enumitem}
\usepackage{fancyhdr}
\usepackage{titletoc}
\usepackage{etoc}
\usepackage{lastpage}
\usepackage{xcolor}
\usepackage{tikz}
\usepackage{eso-pic}
\usepackage{url}
\usepackage{morefloats}   
\usepackage{placeins}     
\usetikzlibrary{positioning,shapes.geometric}

\setcitestyle{numbers,square}

\newcommand{\fito}{\textsc{fito}}
\newcommand{\sgn}{\operatorname{sgn}}

\tikzset{
  badge/.style={
    circle,
    draw=#1,
    fill=#1!10,
    line width=0.5pt,
    minimum size=0.45in,
    font=\tiny\sffamily,
    align=center,
    text=#1!80!black
  }
}

\newcommand{\placebadges}{%
  \ifarxiv\else
  \AddToShipoutPictureBG*{%
    \AtPageUpperLeft{%
      \raisebox{-0.5in}{\hspace{\dimexpr\paperwidth-0.7in\relax}%
        \begin{tikzpicture}[overlay]
          \node[badge=green!60!black] (b1) {Artifacts\\Available};
          \node[badge=purple!70!black, left=0.08in of b1]   (b2) {Expert\\Verified};
          \node[badge=green!50!black,  left=0.08in of b2]   (b3) {AI\\Assisted};
          \node[badge=blue!70!black,   left=0.08in of b3]   (b4) {Human\\Conceived};
        \end{tikzpicture}%
      }%
    }%
  }%
  \fi
}

\newcommand{\maketrcover}{%
  \ifarxiv\else
  \thispagestyle{empty}
  \begin{fullwidth}
  \vspace*{2in}
  \begin{center}
    {\Large\sffamily\bfseries D\AE D\AE LUS Technical Report}\\[1.5em]
    {\LARGE\sffamily\bfseries The Markovianity of Time:\\[0.3em]
     The Category Mistake in Open Quantum Systems}\\[2em]
    {\large \trauthor\,,\;\traffiliation}\\[1em]
    {\normalsize \trversion\quad---\quad\trdate}
  \end{center}

  \vspace{2em}
  \noindent\rule{\linewidth}{0.4pt}

  \vspace{1em}
  \footnotesize
  \begin{description}[leftmargin=1.2in, style=sameline, font=\normalfont\scshape]
    \item[Status:]       v0.1 --- first draft
    \item[Filename:]     \texttt{\trfilename}
    \item[Keywords:]     Markov approximation, arrow of time, time symmetry,
                         FITO, category mistake, Lindblad equation,
                         Caldeira--Leggett, open quantum systems,
                         distributed computing
    \item[Related:]      Lamport's Arrow of Time (arXiv:2602.21730),
                         Circumventing FLP (arXiv, 2026),
                         Semantic Arrow of Time Part~I (arXiv:2603.01440)
    \item[License:]      \textcopyright\ 2026 \trauthor, \traffiliation.
                         All rights reserved.
  \end{description}

  \vspace{1.5em}
  \noindent\rule{\linewidth}{0.4pt}

  \vspace{2em}
  \begin{center}
    \normalsize\itshape
    This cover page may be discarded when printing.\\
    The paper begins on the following page.
  \end{center}

  \end{fullwidth}
  \clearpage
  \fi
}

\fancypagestyle{plain}{%
  \fancyhf{}%
  \fancyfoot[L]{\small\textsc{\trshorttitle}}%
  \fancyfoot[C]{\small\thepage\ of \pageref{LastPage}}%
  \ifarxiv\else
  \fancyfoot[R]{\scriptsize\texttt{\trfilename}}%
  \fi
  \ifarxiv\else
  \fancyfoot[L]{\raisebox{-0.8in}{\tiny\texttt{\trversion\ -- \trdate}}}%
  \fancyfoot[R]{\raisebox{-0.8in}{\tiny\texttt{\trfilename}}}%
  \fi
}
\title{The Markovianity of Time: The Category Mistake in Open Quantum Systems}
\author[Paul Borrill]{Paul Borrill, D\AE D\AE LUS}
\date{02026-MAR-13}

\begin{document}
\maketrcover
\setcounter{page}{1}
\maketitle
\placebadges
\thispagestyle{plain}

\daemargintoc

\begin{abstract}
\noindent The Markov approximation is arguably the most ubiquitous tool in physics, underpinning quantum master equations, stochastic processes, and---via Shannon's channel model and Lamport's logical clocks---the foundational assumptions of distributed computing.
It is widely assumed that Markovianity inherently implies temporal asymmetry: that the Markov property is a forward-in-time-only (\fito{}) construct.
We show that this assumption is a \emph{category mistake} in the sense of Ryle~\citep{ryle1949}.

Guff, Shastry, and Rocco~\citep{guff2025} have recently demonstrated that the Markov approximation applied to the Caldeira--Leggett model---a paradigmatic open quantum system---maintains time-reversal symmetry in the derived equations of motion.
The resulting time-symmetric formulations of quantum Brownian motion, Lindblad master equations, and Pauli master equations describe thermalisation that can occur in two opposing temporal directions.
Asymmetry arises not from the dynamics but from boundary conditions.

We trace how Markovianity's assumed directionality propagated from physics through Shannon's information theory to Lamport's happens-before relation and the impossibility theorems of distributed computing (FLP, CAP, Two Generals).
Each step encodes \fito{} as convention, then treats it as physical law---the same category mistake repeated across domains.
The Surrey result establishes that this conflation is not merely philosophically suspect but \emph{mathematically unnecessary}: the most fundamental approximation used to derive irreversibility is itself time-symmetric.
\end{abstract}

\FloatBarrier
\section[Introduction]{Introduction}
\label{sec:intro}

The Markov property---the principle that the future of a stochastic process depends only on the present, not on the past---is one of the most widely applied mathematical assumptions in science and engineering.%
\marginalia{The Markov property appears in quantum master equations, chemical kinetics, financial models, queuing theory, population genetics, speech recognition, and virtually every Monte Carlo simulation in computational physics. Its reach is extraordinary.}
In physics, the Born--Markov approximation underpins every standard derivation of dissipative dynamics in open quantum systems.
In information theory, Shannon's noisy channel theorem assumes memoryless noise.
In distributed computing, Lamport's happens-before relation and the consensus protocols it spawned encode a Markovian causal structure in which the ordering of events depends only on the most recent message exchange.

Across all these domains, a common assumption operates silently: that Markovianity is inherently \emph{directional}---that the Markov property implies, or at minimum requires, a distinguished forward arrow of time.
The present paper demonstrates that this assumption is a category mistake.

The evidence comes from a recent paper by Guff, Shastry, and Rocco at the University of Surrey~\citep{guff2025}, published in \emph{Scientific Reports} in January 2025.
Working with the Caldeira--Leggett model---the paradigmatic description of a quantum particle coupled to an infinite heat bath---they show that the Markov approximation does \emph{not} break time-reversal symmetry.
The standard derivation introduces temporal asymmetry through an asymmetric implementation of the limiting procedure, not through the mathematics itself.
When the Markov approximation is performed without this asymmetric truncation, the resulting equations of motion are time-symmetric: dissipation occurs toward equilibrium in \emph{both} temporal directions, with a sign function $\sgn(t)$ governing the direction of the friction term.

This result has profound implications.
If the most fundamental and ubiquitous approximation used to derive irreversibility is itself time-symmetric, then the entire chain of reasoning by which temporal directionality enters physics, information theory, and computing must be re-examined.

We trace this chain through four links:
\begin{enumerate}
    \item The derivation of quantum master equations from the Born--Markov approximation (physics).
    \item Shannon's unidirectional channel model and its encoding of \fito{} as information-theoretic convention (information theory).
    \item Lamport's happens-before relation and its elevation of message-ordering convention to causal law (distributed computing).
    \item The impossibility theorems---FLP, CAP, Two Generals---which are consequences of the \fito{} assumption, not fundamental physical constraints (theoretical computer science).
\end{enumerate}
At each link, the same category mistake occurs: a mathematical convention (forward-time integration, unidirectional channels, acyclic event ordering) is mistaken for a physical necessity.

The Category Mistake framework, developed in earlier papers in this series~\citep{borrill2026lamport,borrill2026flp,borrill2026cap}, identifies this error as an instance of the \emph{Forward-In-Time-Only} (\fito{}) assumption---the commitment that information flows, causation operates, and system evolution proceeds exclusively from past to future.
The Surrey paper provides the most fundamental evidence yet that \fito{} is optional, not obligatory: it is a boundary condition, not a law of dynamics.

\FloatBarrier
\section[The Surrey Result]{The Surrey Result: Time-Symmetric Markovianity}
\label{sec:surrey}

Guff, Shastry, and Rocco~\citep{guff2025} begin with the standard Caldeira--Leggett Hamiltonian:
a system $S$ coupled bilinearly to an infinite bath $B$ of harmonic oscillators.
The total Hamiltonian is $H = H_S + H_B + H_I$, where $H_I$ represents the system--bath interaction.

The standard procedure traces out the bath degrees of freedom using the Born approximation (weak coupling, factorised initial state $\rho(t_0) \approx \rho_S(t_0) \otimes \rho_B$) and the Markov approximation (bath correlations decay much faster than the system dynamics, allowing integration limits to be extended to infinity).%
\marginalia{The Caldeira--Leggett model (1983) is the benchmark for open quantum systems. The key insight: textbooks present the Markov limit as requiring $t \to \infty$ (forward only), but the correct implementation uses $|t| \to \infty$---preserving time-reversal symmetry.}

\subsection{Where the Asymmetry Actually Enters}

The key insight of the Surrey paper is that the standard derivation introduces temporal asymmetry not through the Markov approximation itself, but through an \emph{asymmetric implementation} of the limiting procedure.

Specifically, the bath correlation function (the memory kernel) $k(\tau)$ is an even function: $k(\tau) = k(-\tau)$.
When performing the Markov limit, one must evaluate integrals of the form
\begin{equation}
\label{eq:memory}
\int_0^{|t|} d\tau \; k(\tau) \, f(\tau),
\end{equation}
where the integration extends to $|t|$, not to $t$.
The standard textbook treatment silently replaces $|t|$ with $t$, which is valid only for $t > 0$---thereby hardcoding a forward-time bias into what is presented as a ``derivation'' of irreversibility.

\subsection{The Time-Symmetric Equation of Motion}

When the Markov approximation is performed correctly, respecting the even symmetry of $k(\tau)$, the resulting quantum Langevin equation for the system coordinate $\hat{Q}$ takes the form:
\begin{equation}
\label{eq:langevin}
M \frac{d^2\hat{Q}}{dt^2} + V'(\hat{Q}(t)) + \sgn(t)\,\gamma\,\hat{P}(t) = \hat{f}(t),
\end{equation}
where $\gamma$ is the damping coefficient, $\hat{P}$ is the momentum operator, and $\hat{f}(t)$ is the fluctuating force from the bath.%
\marginalia{The crucial term is $\sgn(t)\,\gamma\,\hat{P}(t)$. For $t > 0$, dissipation drives the system toward equilibrium in the forward direction. For $t < 0$, dissipation drives toward equilibrium in the \emph{backward} direction. Both arrows of time emerge from the \emph{same equation}.}

The sign function $\sgn(t)$ is the signature of time-symmetric dissipation.
For $t > 0$, the friction term $+\gamma\hat{P}$ drives the system toward thermal equilibrium in the forward-time direction.
For $t < 0$, the term $-\gamma\hat{P}$ drives toward equilibrium in the backward-time direction.
The equation of motion is symmetric under time reversal $t \to -t$; what breaks is time-\emph{translation} invariance (the semigroup property) at $t = 0$, not time-\emph{reversal} symmetry.

\subsection{Propagation to Master Equations}

The time-symmetric structure propagates to the three fundamental quantum master equations:
\begin{enumerate}
    \item \textbf{Quantum Brownian motion:} The Fokker--Planck equation derived from (\ref{eq:langevin}) inherits the $\sgn(t)$ factor, yielding diffusion toward equilibrium in both temporal directions.
    \item \textbf{Lindblad (GKSL) master equation:} The Lindblad dissipator $\mathcal{D}[\rho]$, which governs decoherence and dissipation in the density matrix formalism, can be derived with time-reversal symmetry intact. The dissipative superoperator does not inherently break $T$-symmetry.%
    \marginalia{The Lindblad equation~\citep{lindblad1976,gorini1976} is $\dot{\rho} = -i[H,\rho] + \sum_k (L_k\rho L_k^\dagger - \frac{1}{2}\{L_k^\dagger L_k, \rho\})$. Guff et al.\ show that the Lindblad operators $L_k$ satisfy time-reversal symmetry when derived from the Caldeira--Leggett model with the correct Markov limit.}
    \item \textbf{Pauli master equation:} The rate equations governing transitions between energy eigenstates maintain detailed balance in both temporal directions, consistent with the symmetric and antisymmetric forms identified by Tay and Petrosky~\citep{tay2006}.
\end{enumerate}

The resulting evolution map takes the form:
\begin{equation}
\label{eq:evolution}
\hat{\mathcal{E}}(t) = \exp\!\bigl(i\hat{\mathcal{L}}_H t + \hat{\mathcal{L}}_D |t|\bigr),
\end{equation}
where $\hat{\mathcal{L}}_H$ is the Hamiltonian (unitary) part and $\hat{\mathcal{L}}_D$ is the dissipative part.
The absolute value $|t|$ ensures that dissipation drives toward equilibrium regardless of temporal direction.%
\marginalia[-2cm]{This is a Markov \emph{process group}---it acts as a semigroup for both $t > 0$ and $t < 0$, but is not a semigroup across $t = 0$. The initial condition at $t = 0$ is the sole source of apparent temporal asymmetry.}

\subsection{The Origin of the Arrow}

If the dynamics are time-symmetric, where does the arrow of time come from?
The answer is the same one that Boltzmann identified in the nineteenth century and that Price~\citep{price1996} has forcefully articulated in the twentieth: \emph{boundary conditions}.%
\marginalia{Boltzmann's $H$-theorem assumes molecular chaos (\emph{Stosszahlansatz})---an initial condition, not a dynamical law. The Surrey result shows that the quantum analogue (the Born--Markov approximation) has the same character: the approximation itself is time-symmetric; the arrow comes from the initial state.}

A low-entropy initial state at $t = 0$ constrains the system to evolve toward higher entropy for $t > 0$.
But the \emph{same dynamics}, applied with a low-entropy ``final'' state, would produce increasing entropy for $t < 0$.
The arrow of time is selected by the boundary condition, not by the equations of motion.

This is precisely the point that Huw Price~\citep{price1996} makes at the macroscopic level and that Aharonov's two-state vector formalism~\citep{aharonov1964} makes at the quantum level: if you impose boundary conditions at both ends, time-symmetric dynamics naturally generate what \emph{appears} to be a directed process from the perspective of an observer embedded within the evolution.%
\marginalia{Ryle's category mistake: a visitor to Oxford, shown the colleges and playing fields, asks ``But where is the University?''---confusing an abstraction with a physical place. The \fito{} error is the temporal analogue: confusing a boundary condition with a dynamical law.}

\FloatBarrier
\section[The Category Mistake]{The Category Mistake: Confusing Convention with Dynamics}
\label{sec:mistake}

Gilbert Ryle introduced the concept of a \emph{category mistake} to describe the error of presenting a fact or entity as if it belonged to a logical category other than its own~\citep{ryle1949}.
We identify a category mistake in the standard treatment of Markovianity that operates at three levels:

\subsection{Level 1: Physics}

The standard derivation of quantum master equations treats the Markov approximation as if it \emph{produces} irreversibility.
Textbooks present the chain: Born approximation $\to$ Markov approximation $\to$ Lindblad equation $\to$ irreversible dissipation $\to$ arrow of time.%
\marginalia{See, for instance, Breuer and Petruccione~\citep{breuer2002}, Chapter~3. The standard derivation integrates only forward in time. The Surrey paper shows this is a convention, not a necessity.}

But as the Surrey result demonstrates, the Markov approximation is time-symmetric.
The irreversibility is introduced by the \emph{boundary condition} (the factorised initial state at $t_0$), not by the approximation.
Attributing irreversibility to the Markov property is a category mistake: it confuses a \emph{mathematical technique} (the memoryless approximation) with a \emph{physical claim} (the direction of time).

\begin{center}
\begin{tabular}{@{}lll@{}}
\toprule
\textbf{Category} & \textbf{Correct attribution} & \textbf{Category mistake} \\
\midrule
Markov property & memoryless dynamics & arrow of time \\
Born--Markov limit & $|t| \to \infty$ & $t \to +\infty$ \\
Dissipation & toward equilibrium (either direction) & forward-only decay \\
Initial condition & boundary constraint & dynamical law \\
\bottomrule
\end{tabular}
\end{center}

\subsection{Level 2: Information Theory}

Shannon's noisy channel model~\citep{shannon1948} encodes the Markov property as forward-only information flow: Source $\to$ Encoder $\to$ Channel $\to$ Decoder $\to$ Destination.
The channel noise is memoryless (Markovian), and information flows exclusively from source to destination.%
\marginalia[-2cm]{Shannon himself was likely aware that this was a modeling choice. His mathematical framework is symmetric: mutual information satisfies $I(X;Y) = I(Y;X)$, regardless of which variable is ``sent'' and which ``received.'' The directionality is in the \emph{model}, not in the \emph{mathematics}.}

Yet the mathematics of information is symmetric.
Mutual information satisfies $I(X;Y) = I(Y;X)$---the same symmetry that appears in the Surrey result.
Shannon's channel model is the information-theoretic analogue of the asymmetric Markov limit: it takes a symmetric mathematical object and imposes directionality by convention.

The category mistake: treating the directionality of Shannon's \emph{channel model} as if it were a property of \emph{information itself}.

\subsection{Level 3: Distributed Computing}

Lamport's happens-before relation~\citep{lamport1978} completes the chain.
Rule~(2)---that sending precedes receiving---encodes Shannon's unidirectional channel as a causal axiom.
The Markov property, now twice removed from its mathematical origins, appears as the assumption that the state of a distributed system depends only on the most recent messages, and that these messages flow in one direction.%
\marginalia[-1cm]{Lamport's Rule~(2) is the exact analogue of the textbook Markov limit's silent replacement of $|t|$ with $t$. Both introduce directionality into a structure that does not require it. Both are conventions presented as derivations.}

The happens-before relation embeds this directional Markov structure into a DAG---a directed acyclic graph.
The impossibility theorems (FLP~\citep{flp1985}, CAP~\citep{brewer2000,gilbert2002}, Two Generals) are then proved within this DAG framework.

The category mistake: treating the impossibility of consensus \emph{within a FITO model} as if it were an impossibility of consensus \emph{in nature}.

\subsection{The Chain of Conflation}

The full chain of the category mistake is:
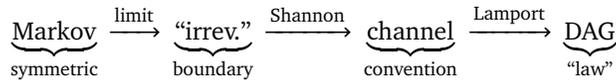
\begin{equation*}
\underbrace{\text{Markov}}_{\text{symmetric}}
\;\xrightarrow{\text{limit}}\;
\underbrace{\text{``irrev.''}}_{\text{boundary}}
\;\xrightarrow{\text{Shannon}}\;
\underbrace{\text{channel}}_{\text{convention}}
\;\xrightarrow{\text{Lamport}}\;
\underbrace{\text{DAG}}_{\text{``law''}}
\end{equation*}
At each step, a symmetric mathematical object acquires a directionality that is then treated as fundamental by the next stage.
The Surrey result breaks the chain at its root: the Markov property is time-symmetric, and everything downstream that depends on its alleged asymmetry is built on a category mistake.

\FloatBarrier
\section[From Physics to Computing]{From Physics to Computing: The Propagation of \fito{}}
\label{sec:propagation}

Having identified the category mistake at its mathematical origin, we now trace its propagation through the intellectual history that connects quantum dissipation to distributed systems theory.

\subsection{Markov Chains in Computing}

Markov chains are fundamental to computing far beyond distributed systems.
They underpin Monte Carlo simulation, PageRank~\citep{brin1998}, speech recognition, reinforcement learning, and probabilistic model checking.%
\marginalia{PageRank models web surfing as a Markov chain on the link graph. The directedness is a property of \emph{the web}, not of Markov chains. Indeed, the time-reversal of any stationary chain with transition matrix~$P$ and stationary distribution~$\pi$ is itself Markovian, with $\tilde{P}_{ij} = \pi_j P_{ji}/\pi_i$~\citep{kelly1979}.}
In every application, the convention is the same: the chain transitions \emph{forward} through a sequence of states $X_0, X_1, X_2, \ldots$

But the Markov property itself---$P(X_{n+1} | X_n, X_{n-1}, \ldots) = P(X_{n+1} | X_n)$---says nothing about temporal direction.
It states that the conditional distribution of the next state depends only on the current state.
The same property holds for ``previous'' states: a time-reversed Markov chain is still Markovian.

The Surrey result is the quantum-mechanical version of this classical fact: Markov dynamics in quantum systems are time-symmetric.
The forward convention is just that---a convention.%
\marginalia{The FLP impossibility construction depends on the assumption that processes cannot ``reach back'' to determine a peer's state---exactly the \fito{} assumption at the protocol level.}

\subsection{Consensus and the Forward Assumption}

The FLP impossibility result~\citep{flp1985} proves that no deterministic asynchronous protocol can guarantee consensus with even one crash failure.
The proof exploits the impossibility of distinguishing, from any single process's perspective, a crashed process from a slow one.

This indistinguishability is a consequence of \fito{}: information can only flow forward in time, so a process can only wait for messages that have not yet arrived.
It cannot inspect the state of the sender at the moment of (non-)sending.

The CAP theorem~\citep{brewer2000,gilbert2002} similarly assumes that consistency requires maintaining a total order on operations---a global DAG of events.
Under network partitions, the \fito{} assumption prevents nodes from establishing this order, forcing a choice between consistency and availability.

Both results are theorems about \fito{} systems.
Neither addresses whether non-\fito{} coordination---bilateral, time-symmetric information exchange---might evade the impossibility.%
\marginalia{This is the constructive programme of the Category Mistake series: to demonstrate that bilateral coordination mechanisms (e.g., bisynchronous FIFOs~\citep{borrill2026bisync}) can achieve what \fito{} protocols cannot. The present paper establishes the physical basis for this claim.}

\subsection{The Wheeler--Feynman Absorber Theory and Bilateral Channels}

The physics of time-symmetric communication has been explored since Wheeler and Feynman's absorber theory~\citep{wheeler1945}, which showed that classical electrodynamics admits both retarded (forward) and advanced (backward) solutions on equal footing.%
\marginalia[-0.5cm]{Wheeler and Feynman showed that the apparent dominance of retarded radiation is a consequence of the \emph{absorber} boundary condition---an infinite future absorber that cancels advanced waves. The asymmetry is in the boundary, not in Maxwell's equations. This is the electromagnetic precursor of the Surrey result.}
Cramer's transactional interpretation~\citep{cramer1986} extends this to quantum mechanics, modeling quantum events as bilateral ``handshakes'' between retarded offer waves and advanced confirmation waves.

In the language of distributed computing, Wheeler--Feynman describes a \emph{bilateral channel}: information flows in both temporal directions, and the ``transaction'' (the observed radiation) is the symmetric result of both flows.
This is precisely the model that \fito{} excludes.

\FloatBarrier
\section[Implications]{Implications}
\label{sec:implications}

The Surrey result, read through the lens of the Category Mistake framework, has implications across multiple domains.

\subsection{For Physics}

The time-symmetric Markov formulation resolves a long-standing puzzle: how can time-symmetric microscopic dynamics give rise to macroscopic irreversibility?%
\marginalia{This is the Loschmidt paradox, raised in 1876 against Boltzmann's $H$-theorem. Boltzmann's answer (low-entropy initial conditions) is correct but has been obscured by the conflation of the Markov approximation with irreversibility. The Surrey result restores Boltzmann's answer to its proper form.}
The answer is not that the dynamics secretly break time symmetry; it is that boundary conditions select a direction.
The Markov approximation, Lindblad equation, and Pauli master equation are all \emph{compatible} with both arrows of time.
What selects the observed arrow is the low-entropy initial state of the universe---a cosmological fact, not a dynamical one.

This coheres with the time-symmetric interpretations of quantum mechanics developed by Aharonov~\citep{aharonov1964}, Price~\citep{price1996}, and Wharton~\citep{wharton2010}, and with the experimental demonstrations of indefinite causal order~\citep{rubino2017,goswami2018}.
The process matrix formalism of Oreshkov, Costa, and Brukner~\citep{oreshkov2012} further demonstrates that quantum mechanics does not require a definite causal order---let alone a forward-only one.

\subsection{For Distributed Computing}

If the Markov property does not require temporal asymmetry, then the mathematical foundations on which Lamport's happens-before relation, FLP, and CAP rest are weaker than generally assumed.%
\marginalia[-2cm]{We emphasise: the impossibility theorems are not \emph{wrong}. They are theorems about a specific model. The question is whether the model is the only possible one.}

Specifically:
\begin{enumerate}
    \item \textbf{Lamport clocks are coordinate systems.} Just as general relativity shows that coordinate systems are representational, not physical, logical clocks represent a \emph{choice of embedding}, not the causal structure itself. Alternative embeddings---bilateral, symmetric, non-DAG---are mathematically valid.
    \item \textbf{FLP is model-specific.} The impossibility of consensus under one crash failure is a theorem about \fito{} protocols in asynchronous systems. Bilateral coordination mechanisms may achieve consensus by different means.
    \item \textbf{CAP can be circumvented.} The consistency--availability--partition tolerance trade-off assumes that consistency requires a total order (a DAG). If coordination is based on mutual information conservation rather than temporal ordering, the trade-off may not apply in its present form.
\end{enumerate}

\subsection{For Engineering Practice}

The practical engineering implication is that distributed systems need not be designed exclusively around forward-only message passing.%
\marginalia{Bisynchronous FIFOs~\citep{borrill2026bisync} provide a silicon-proven example: bilateral, non-\fito{} coordination at the hardware level. The Category Mistake series argues that this principle extends to software protocols.}
Bilateral coordination---in which both endpoints contribute symmetrically to the establishment of shared state---is not merely a theoretical curiosity.
It is the default mode of physical interaction, from which \fito{} is a deliberate (and limiting) departure.

\FloatBarrier
\section[Discussion]{Discussion}
\label{sec:discussion}

The Surrey result is not, in itself, a surprise to physicists familiar with time-symmetric formulations of quantum mechanics.
What is surprising is how completely the \fito{} reading of Markovianity has been accepted as canonical, and how far-reaching its consequences have been.%
\marginalia[-1cm]{The acceptance is so complete that the Surrey paper itself received significant media attention as a ``discovery''---despite the mathematical ingredients being available for decades. The news was not the time symmetry of the Markov approximation, but the fact that it had been overlooked.}

The standard textbook treatment of open quantum systems~\citep{breuer2002} integrates forward in time as a matter of course.
Shannon's channel model sends information from left to right.
Lamport's Rule~(2) points the arrow from sender to receiver.
At no point does anyone announce ``we are now assuming \fito{}.''
The assumption enters silently, disguised as mathematical convenience, and accumulates the authority of physical law through repetition.

This is the anatomy of a category mistake: not a dramatic error, but a quiet drift of ontological status.
A convention becomes a derivation.
A derivation becomes a theorem.
A theorem becomes a law.
And a law forecloses the design space.

The contribution of this paper is to identify the specific mathematical point at which the mistake enters (the asymmetric Markov limit), to cite the specific result that demonstrates its unnecessity (the Surrey paper), and to trace the specific path by which it propagated from quantum physics to distributed computing theory.%
\marginalia{The chain: Markov approximation (physics) $\to$ Shannon channel (information theory) $\to$ Lamport Rule~(2) (distributed computing) $\to$ FLP/CAP/Two Generals (impossibility theorems). Each step inherits the asymmetry from its predecessor without re-examining it.}

\FloatBarrier
\section[Conclusion]{Conclusion}
\label{sec:conclusion}

The Markov approximation is time-symmetric.
This is not a new philosophical claim but a mathematical fact, demonstrated rigorously by Guff, Shastry, and Rocco~\citep{guff2025} for the Caldeira--Leggett model and its three most important descendants: the quantum Brownian motion equation, the Lindblad master equation, and the Pauli master equation.

The assumption that Markovianity implies a forward arrow of time is a category mistake---a conflation of mathematical convention with physical dynamics.
This mistake propagated from physics through Shannon's information theory to Lamport's distributed computing foundations, where it acquired the status of physical law and produced the impossibility theorems (FLP, CAP, Two Generals) that define the boundaries of the field.

Those boundaries are real, but they are boundaries of a \emph{model}, not of nature.
The model assumes \fito{}.
Nature does not.%
\marginalia{``The model assumes \fito{}. Nature does not.'' This sentence summarises the entire Category Mistake series.}

The Markov property tells us that dynamics are memoryless.
It does not tell us which direction to forget.

\section*{Acknowledgments}

The author acknowledges the use of AI-assisted tools (Anthropic Claude) for literature review and structural refinement.
All interpretations, claims, and conclusions are solely the responsibility of the author.
This work is part of a 25-year research programme spanning VERITAS, REPLICUS, Earth Computing, and D\AE D\AE LUS.


\bibliographystyle{unsrtnat}
\bibliography{references}

\begin{thebibliography}{25}
\providecommand{\natexlab}[1]{#1}
\providecommand{\url}[1]{\texttt{#1}}
\expandafter\ifx\csname urlstyle\endcsname\relax
  \providecommand{\doi}[1]{doi: #1}\else
  \providecommand{\doi}{doi: \begingroup \urlstyle{rm}\Url}\fi

\bibitem[Ryle(1949)]{ryle1949}
Gilbert Ryle.
\newblock \emph{The Concept of Mind}.
\newblock Hutchinson, London, 1949.

\bibitem[Guff et~al.(2025)Guff, Shastry, and Rocco]{guff2025}
Thomas Guff, Chintalpati~Umashankar Shastry, and Andrea Rocco.
\newblock Emergence of opposing arrows of time in open quantum systems.
\newblock \emph{Scientific Reports}, 15:\penalty0 3658, 2025.
\newblock \doi{10.1038/s41598-025-87323-x}.
\newblock arXiv:2311.08486.

\bibitem[Borrill(2026{\natexlab{a}})]{borrill2026lamport}
Paul Borrill.
\newblock Lamport's arrow of time: The category mistake in logical clocks,
  2026{\natexlab{a}}.
\newblock arXiv:2602.21730 [cs.DC].

\bibitem[Borrill(2026{\natexlab{b}})]{borrill2026flp}
Paul Borrill.
\newblock Circumventing {FLP}: The category mistake in asynchronous consensus,
  2026{\natexlab{b}}.
\newblock arXiv preprint.

\bibitem[Borrill(2026{\natexlab{c}})]{borrill2026cap}
Paul Borrill.
\newblock Circumventing {CAP}: The category mistake in distributed consistency,
  2026{\natexlab{c}}.
\newblock arXiv preprint.

\bibitem[Lindblad(1976)]{lindblad1976}
G{\"o}ran Lindblad.
\newblock On the generators of quantum dynamical semigroups.
\newblock \emph{Communications in Mathematical Physics}, 48:\penalty0 119--130,
  1976.
\newblock \doi{10.1007/BF01608499}.

\bibitem[Gorini et~al.(1976)Gorini, Kossakowski, and Sudarshan]{gorini1976}
Vittorio Gorini, Andrzej Kossakowski, and E.~C.~G. Sudarshan.
\newblock Completely positive dynamical semigroups of {N}-level systems.
\newblock \emph{Journal of Mathematical Physics}, 17:\penalty0 821--825, 1976.
\newblock \doi{10.1063/1.522979}.

\bibitem[Tay and Petrosky(2016)]{tay2006}
Boon~Leong Tay and Tomio Petrosky.
\newblock Symmetric and antisymmetric forms of the {Pauli} master equation.
\newblock \emph{Scientific Reports}, 6:\penalty0 29942, 2016.
\newblock \doi{10.1038/srep29942}.

\bibitem[Price(1996)]{price1996}
Huw Price.
\newblock \emph{Time's Arrow and {Archimedes}' Point: New Directions for the
  Physics of Time}.
\newblock Oxford University Press, 1996.

\bibitem[Aharonov et~al.(1964)Aharonov, Bergmann, and Lebowitz]{aharonov1964}
Yakir Aharonov, Peter~G. Bergmann, and Joel~L. Lebowitz.
\newblock Time symmetry in the quantum process of measurement.
\newblock \emph{Physical Review}, 134:\penalty0 B1410--B1416, 1964.
\newblock \doi{10.1103/PhysRev.134.B1410}.

\bibitem[Breuer and Petruccione(2002)]{breuer2002}
Heinz-Peter Breuer and Francesco Petruccione.
\newblock \emph{The Theory of Open Quantum Systems}.
\newblock Oxford University Press, 2002.

\bibitem[Shannon(1948)]{shannon1948}
Claude~E. Shannon.
\newblock A mathematical theory of communication.
\newblock \emph{Bell System Technical Journal}, 27:\penalty0 379--423,
  623--656, 1948.

\bibitem[Lamport(1978)]{lamport1978}
Leslie Lamport.
\newblock Time, clocks, and the ordering of events in a distributed system.
\newblock \emph{Communications of the ACM}, 21\penalty0 (7):\penalty0 558--565,
  1978.
\newblock \doi{10.1145/359545.359563}.

\bibitem[Fischer et~al.(1985)Fischer, Lynch, and Paterson]{flp1985}
Michael~J. Fischer, Nancy~A. Lynch, and Michael~S. Paterson.
\newblock Impossibility of distributed consensus with one faulty process.
\newblock \emph{Journal of the ACM}, 32\penalty0 (2):\penalty0 374--382, 1985.
\newblock \doi{10.1145/3149.214121}.

\bibitem[Brewer(2000)]{brewer2000}
Eric~A. Brewer.
\newblock Towards robust distributed systems.
\newblock In \emph{Proceedings of the 19th ACM Symposium on Principles of
  Distributed Computing (PODC)}, 2000.
\newblock Keynote address.

\bibitem[Gilbert and Lynch(2002)]{gilbert2002}
Seth Gilbert and Nancy Lynch.
\newblock Brewer's conjecture and the feasibility of consistent, available,
  partition-tolerant web services.
\newblock \emph{ACM SIGACT News}, 33\penalty0 (2):\penalty0 51--59, 2002.
\newblock \doi{10.1145/564585.564601}.

\bibitem[Brin and Page(1998)]{brin1998}
Sergey Brin and Lawrence Page.
\newblock The anatomy of a large-scale hypertextual web search engine.
\newblock \emph{Computer Networks and ISDN Systems}, 30:\penalty0 107--117,
  1998.
\newblock \doi{10.1016/S0169-7552(98)00110-X}.

\bibitem[Kelly(1979)]{kelly1979}
Frank~P. Kelly.
\newblock \emph{Reversibility and Stochastic Networks}.
\newblock John Wiley \& Sons, 1979.

\bibitem[Borrill(2026{\natexlab{d}})]{borrill2026bisync}
Paul Borrill.
\newblock Bisynchronous {FIFOs} and the {FITO} category mistake,
  2026{\natexlab{d}}.
\newblock arXiv:2603.03470 [cs.DC].

\bibitem[Wheeler and Feynman(1945)]{wheeler1945}
John~A. Wheeler and Richard~P. Feynman.
\newblock Interaction with the absorber as the mechanism of radiation.
\newblock \emph{Reviews of Modern Physics}, 17:\penalty0 157--181, 1945.
\newblock \doi{10.1103/RevModPhys.17.157}.

\bibitem[Cramer(1986)]{cramer1986}
John~G. Cramer.
\newblock The transactional interpretation of quantum mechanics.
\newblock \emph{Reviews of Modern Physics}, 58:\penalty0 647--687, 1986.
\newblock \doi{10.1103/RevModPhys.58.647}.

\bibitem[Wharton(2010)]{wharton2010}
Ken~B. Wharton.
\newblock A novel interpretation of the {Klein--Gordon} equation.
\newblock \emph{Foundations of Physics}, 40:\penalty0 313--332, 2010.
\newblock \doi{10.1007/s10701-009-9398-2}.

\bibitem[Rubino et~al.(2017)Rubino, Rozema, Feix, Ara{\'u}jo, Zeuner, Procopio,
  Brukner, and Walther]{rubino2017}
Giulia Rubino, Lee~A. Rozema, Adrien Feix, Mateus Ara{\'u}jo, Jonas~M. Zeuner,
  Lorenzo~M. Procopio, {\v{C}}aslav Brukner, and Philip Walther.
\newblock Experimental verification of an indefinite causal order.
\newblock \emph{Science Advances}, 3:\penalty0 e1602589, 2017.
\newblock \doi{10.1126/sciadv.1602589}.

\bibitem[Goswami et~al.(2018)Goswami, Giarmatzi, Kewming, Costa, Branciard,
  Romero, and White]{goswami2018}
Kejin Goswami, Christina Giarmatzi, Michael Kewming, Fabio Costa, Cyril
  Branciard, Jacquiline Romero, and Andrew~G. White.
\newblock Indefinite causal order in a quantum switch.
\newblock \emph{Physical Review Letters}, 121:\penalty0 090503, 2018.
\newblock \doi{10.1103/PhysRevLett.121.090503}.

\bibitem[Oreshkov et~al.(2012)Oreshkov, Costa, and Brukner]{oreshkov2012}
Ognyan Oreshkov, Fabio Costa, and {\v{C}}aslav Brukner.
\newblock Quantum correlations with no causal order.
\newblock \emph{Nature Communications}, 3:\penalty0 1092, 2012.
\newblock \doi{10.1038/ncomms2076}.

\end{thebibliography}

\end{document}